\documentclass[journal=jacsat,manuscript=article]{achemso}
\SectionNumbersOn
\usepackage[table]{xcolor}
\usepackage{comment}
\usepackage{hyperref}
\usepackage[version=3]{mhchem} % Formula subscripts using \ce{}
\usepackage{amsmath}

\author{Cecilia Vona}
\author{Mathias Dankl}
\author{Ariadni Boziki}
\author{Martin P. Bircher}
\author{Ursula Rothlisberger}
\email{ursula.roethlisberger@epfl.ch}
\phone{+41 (0)21 6930321}
\fax{+41 (0)21 930320}
\affiliation[Unknown University]
{Laboratory of Computational Chemistry and Biochemistry, Ecole Polytechnique F\'{e}d\'{e}rale de Lausanne (EPFL), CH-1015 Lausanne, Switzerland}

\title[An \textsf{achemso} demo]
  {Force-Matching Based Polarizable and Non-Polarizable Force Fields for Perovskite and Non-Perovskite Phases of CsPbI$_3$}

\abbreviations{IR,NMR,UV}
\keywords{American Chemical Society, \LaTeX}

\begin{document}

\begin{abstract}

  Lead halide perovskites have emerged as highly efficient solar cell materials. However, to date, the most promising members of this class are polymorphs, in which a wide-band gap $\delta$ phase competes with the photoactive perovskite $\alpha$ form, and the intrinsic physical interactions that stabilize one phase over the other are currently not well understood. Classical molecular dynamics simulations based on suitably parameterized force fields (FF) enable computational studies over broad temperature (and pressure) ranges and can help to identify the underlying factors that govern relative phase stability at the atomic level. In this article, we present a polarizable ({\it pol}) as well as a non-polarizable ({\it npol}) FF for the all-inorganic lead halide material CsPbI$_3$ as a prototype system exhibiting a $\delta$/$\alpha$ polymorphism.  Both {\it npol} and {\it pol} FFs have been determined using a force-matching approach based on extensive reference data from first-principles molecular dynamics simulations over a wide range of temperatures. While both FFs are able to describe the perovskite as well as the non-perovskite $\delta$ phase, finer structural details as well as the relative phase stability are better reproduced with the polarizable version. A comparison of these ab initio-derived interatomic potentials allows direct insights into the physical origin of the interactions that govern the interplay between the two competing phases. It turns out that  explicit polarization  is the essential factor that stabilizes the strongly anisotropic $\delta$ phase over  the high symmetry (cubic) perovskite $\alpha$ phase at lower temperatures.  This fundamental difference between  $\alpha$ and $\delta$ phase appears universal and might thus also hold for other perovskite compounds with $\delta$/$\alpha$ polymorphism and thus provide rational guidance for synthetic efforts to stabilize the photoactive perovskite phase at room temperature. 
\end{abstract}

\section{Introduction}
During the last decade, metal halide perovskites have emerged as promising light harvesting materials for next generation solar cells. In fact, the power-conversion efficiencies (PCE) of perovskite solar cells (PSCs) as registered by the national renewable energy laboratory (NREL)\cite{national_renewable_energy_laboratory_best_2019} increased from 3.8\% in 2009 \cite{kojima_organometal_2009} to 26.1\% in 2023 \cite{min_perovskite_2021}. These compounds share the generic stoichiometric formula $\text{ABX}_{3}$ and consist of a divalent metal cation (B = $\text{Pb}^{\text{2+}}$ or $\text{Sn}^{\text{2+}}$), a halide anion (X = $\text{I}^{\text{-}}$, $\text{Br}^{\text{-}}$, $\text{Cl}^{\text{-}}$) and a monovalent cation which can be an organic molecular ion such as A = methylammonium (MA = $\text{CH}_\text{3}\text{NH}_\text{3}^{\text{+}}$), or formamidinium (FA =$ \text{CH(NH}_{\text{2}}\text{)}_{\text{2}}^{\text{+}}$), or an inorganic elemental ion such as A = $\text{Cs}^{+}$. 
The current record efficiencies for single-junction PSCs are based on the perovskite phase of FAPb$\text{I}_{3}$ that features a nearly ideal band gap of 1.44 eV \cite{stoumpos_semiconducting_2013} and exhibits increased thermal stability with respect to  MAPb$\text{I}_{3}$. However, one drawback of FAPb$\text{I}_{3}$ is that the photovoltaically active perovskite $\alpha$ phase is not the most thermodynamically stable form at room temperature, at which the compound assumes a $\delta$ polymorph that is a wide band gap semiconductor unsuitable for use as an efficient solar light absorber. The same type of polymorphism also occurs in the case of the all-inorganic CsPb$\text{I}_{3}$ material, which has become an ideal material for tandem solar cells \cite{ahmad_inorganic_2017,wang_optoelectronic_2022} and light-emitting diodes\cite{sim_phase_2018} thanks to its optical properties along with relatively easy synthesis.
The question of how to control the phase transition of these compounds and stabilize the $\alpha$ phase is an intensely researched topic and several strategies have been devised for this purpose. Among these, there are the use of mixtures of monovalent cations such as FA/MA and FA/Cs mixed compounds, mixed 2D/3D perovskites \cite{saliba_cesium-containing_2016,gelvezrueda_formamidiniumbased_2020,hong_guaninestabilized_2020, mishra_naphthalenediimideformamidinium-based_2021} and the use of additives \cite{alanazi_atomic-level_2019, su_crown_2020,krishna_nanoscale_2021}. However, the preparation of phase-pure and long-term stable $\alpha$ phase remains an issue.\cite{masi_stabilization_2020}. 

In principle, computer simulations could help in deciphering the factors that influence the relative stability and transition between the two phases and thus give important guidance for suitable and rational synthesis approaches. This task can only be achieved by a computational method that can reliably predict the relative finite temperature stability of the competing phases and allow the treatment of sufficiently large sample sizes (and time scales) to avoid artifacts due to finite size effects and/or limited sampling of thermally relevant configurations. These requirements are hard to fulfill at the level of a full first-principles, e.g. density functional theory (DFT) based description, but can more easily be provided when using suitably parameterized force fields (FF). 
Not surprisingly, considerable research efforts have been directed at the development of accurate FFs for lead halide perovskites. 
One of the first FFs (MYP0) was developed by Mattoni et al. for MAPbI$_{\text{3}}$\cite{mattoni_methylammonium_2015}. It adopts three related perovskite phases (orthorhombic, tetragonal and cubic) as a function of temperature while no $\delta$ phase has so far been reported for this system. However, the first-generation MYP0 FF fails to describe the room temperature tetragonal phase. 
MYP0 has been modified \cite{mattoni_modeling_2017} and extended to remedy this shortcoming and to describe the interaction of MAPbI$_{\text{3}}$ with water \cite{caddeo_collective_2017} as well as to capture MAPbBr$_{\text{3}}$ and mixed halide MAPbI$_{3-x}$Br$_x$ compounds \cite{hata_development_2017}. Alternatively, DFT-trained machine-learning potentials (MLPs) for MAPbI$_3$, MAPbBr$_3$ and MAPbCl$_3$ have been generated.\cite{jinnouchi_phase_2019,bokdam_exploring_2021} While the relative stabilities and the transition between different perovskite-like polymorphs can be described fairly well with several of the current FFs, the simultaneous description of the non-perovskite $\delta$ phase and the perovskite phase(s) and the transition between the photovoltaically active and inactive phases, respectively, has remained challenging and the authors are not aware of any attempts made so far to develop such FFs. 

Here, we will focus on the
most simple prototypical compound that exhibits the emergence of a competing $\delta$ phase, the all-inorganic CsPbI$_3$, with the goal of developing accurate ab initio derived FFs that can simultaneously describe both phases and allow insights into the factors that govern relative phase stability. CsPbI$_3$ exhibits a first-order phase transition from the $\delta$ to the cubic $\alpha$ perovskite phase at about 600K \cite{beal_cesium_2016,marronnier_anharmonicity_2018} and upon cooling can be trapped into tetragonal ($\beta$) and orthorhombic ($\gamma$) metastable perovskite phases \cite{stoumpos_renaissance_2015}. 
Apart from its structural properties, the perovskite phase of CsPbI$_3$ (and its mixed halide counterparts) possesses a tunable band gap in the range of 1.72 - 2.3 eV\cite{duan_high-purity_2018}, strong absorption and high photoluminescence quantum yield in the red \cite{lai_direct_2022} which makes it highly suitable for many optoelectronic applications. Not surprisingly, several attempts to develop FFs or MLPs for this system have been undertaken.
A recently developed reactive FF targeted at describing the decomposition of CsPbI$_3$ into PbI$_2$ and CsI focuses on capturing the different perovskite phases of the system; however, it does not take the thermodynamically most stable form at room temperature, i.e. the $\delta$ phase into account \cite{almishal_new_2020}, which is similar to a recent MLP.\cite{wang_interpolating_2021} 
%XXREF: J. Phys. Chem. Lett. 2021, 12, 6070−6077XX. 
A hybrid Embedded Atomic Buckingham - Coulomb (EABC) potential is able to accurately reproduce the density and structural properties of the orthorhombic $\delta$ phase of CsPbI$_3$ but does not seem to yield the transformation to the high-temperature $\alpha$ phase \cite{almishal_new_2020} which has also been reported for the polarizable FF developed by Rathnayake et al. \cite{rathnayake_evaluation_2020}. To the best of our knowledge, so far, no computational approach has successfully described the transition from the $\delta$ to the perovskite phase within a single molecular dynamics simulation. However, a theoretical estimate of the transition temperature has been obtained by a combination of ab initio electronic structure calculations with vibrational entropy contributions modeled via a SchNetPack deep learning potential trained on ab initio MD data. \cite{braeckevelt_accurately_2022}

In view of its simplicity and lower computational effort, it seems worthwhile to fully explore the performance limits of polarizable as well as unpolarizable FFs for this system. Here, we generate optimal nonpolarizable ($npol$) as well as polarizable ($pol$) models using an automated procedure based on a force-matching (FM) approach \cite{ercolessi_interatomic_1994} in which the freely-adjustable parameters of both FF variants are fitted in such a way as to minimize the difference between the forces acting on all atoms with those from first-principles (DFT) based molecular dynamics simulations of both $\alpha$ and $\delta$ phases, covering extended time scales and temperature ranges. By using the same reference data for both models, we can directly assess the additional benefits that can be achieved via a fully polarizable version and gain further insights into the factors that govern the relative phase stability. 

\section{Methods}

\subsection{Nonpolarizable and polarizable interatomic potentials}

The functional form that we have chosen to represent both {\it npol} and {\it pol} versions of the interatomic potential for CsPbI$_3$ is the one of the {\bf A}tomic {\bf M}ultipole {\bf O}ptimized {\bf E}nergetics for {\bf B}iomolecular {\bf A}pplications ({\bf AMOEBA}) FF as implemented in the \texttt{TINKER} package version 7.1 \cite{Ren_AMOEBA_2003,Ponder_Tinker_2010,rackers_tinker_2018}. Originally designed for biomolecular applications, AMOEBA is a next-generation FF with a flexible form and a highly accurate and transferable representation of electrostatic interactions. Metal halide perovskites are known to be soft materials in which polarization effects can play an important role \cite{yan_reversible_2021}. This special property makes them suitable for charge carrier transport\cite{yaffe_local_2017}, but it is also one of the causes of their structural instability.  
Aiming at the development of high-accuracy FFs for these materials, it seems pertinent to include the possibility of polarization effects to assess their potential impact on structural, energetic and dynamic properties.

In $\mathrm{CsPbI}_{3}$, the only contributions to the interatomic potential are from van der Waals (vdW) and  electrostatic interactions:

\begin{equation}
    U(r_{ij})=U_{\mathrm{vdW}}(r_{ij})+U_{\mathrm{el}}(r_{ij}).
\end{equation}

The electrostatic term in the AMOEBA FF includes two components, one describing contributions due to permanent electrostatic interactions and the other due to induced dipoles and quadrupoles, $U_{\mathrm{el}}=U_{\mathrm{el}}^{\mathrm{perm}}+U_{\mathrm{el}}^{\mathrm{ind}}$. $U_{\mathrm{el}}^{\mathrm{perm}}$ describes the electrostatic interactions between atom-centered multipoles $\mathbf{M}=[q,\mu_x,\mu_y,\mu_z,Q_{xx},Q_{xy},Q_{xz},...,Q_{zz}]^{\mathrm{T}}$, where $q$ is the atomic charge, $\boldsymbol{\mu}$ is the (permanent) dipole and $\mathbf{Q}$ the (permanent) quadrupole moment. On the other hand, $U_{\mathrm{el}}^{\mathrm{ind}}$ describes the electronic polarization
that is attributed to the distortions of the electronic density under the influence of an external field originating from the charges in the system. The total dipole and quadrupole moments are the sum of the permanent and the induced contributions, $\boldsymbol{\mu}=\boldsymbol{\mu}^{\mathrm{perm}}+\boldsymbol{\mu}^{\mathrm{ind}}$ and $\mathbf{Q}=\mathbf{Q}^{\mathrm{perm}}+\mathbf{Q}^{\mathrm{ind}}$, respectively.
The electrostatic interaction between two atomic centers $i$ and $j$ is expressed as follows:

\begin{equation}
  U_{\mathrm{el}}(r_{ij})=\mathbf{M}_i^{\mathrm{T}}\mathbf{T}_{ij}\mathbf{M}_{j}. 
\end{equation}
In this expression $r_{ij}$ is the distance between the two atomic centers and $\mathbf{T}_{ij}$ is the multipole interaction matrix:

\begin{equation}
\mathbf{T}_{ij} = 
\begin{pmatrix}
T_{ij} & T_{ij,x} & T_{ij,y} & T_{ij,z} &...\\
T_{ij,x}& T_{ij,xx} & T_{ij,xy} & T_{ij,xz} &...& \\
T_{ij,y} & T_{ij,yx} & T_{ij,yy} & T_{ij,yz} &... \\
T_{ij,z} & T_{ij,zx} & T_{ij,zy} & T_{ij,zz}&...
\\
\vdots & \vdots & \vdots& \vdots & \ddots
\end{pmatrix},
\end{equation}

where the matrix elements are $T_{ij}=(r_{ij})^{-1}$,  $T_{ij,\alpha}=\partial_{\beta}T_{ij}$ and $T_{ij,\beta\gamma}=\partial_{\beta}T_{ij,\beta}$ where $\beta,\gamma=x,y,z$ (in the Supporting Information $\mathbf{T}_{ij}$ is given in an explicit form).
When the interactions involve induced dipoles and quadrupoles they have to be damped to avoid the \textit{polarization catastrophe}, which can occur at very short range~\cite{thole_molecular_1981}. This damping results in the introduction of a smearing function for the charge defined as:
\begin{equation}
\rho=\frac{3 a}{4\pi}\mathrm{e}^{-au^3},
\end{equation}
in which $u={r_{ij}}/(\alpha_i\alpha_j)^{1/6}$ is the effective distance as a function of atomic polarizabilities of sites $i$ and $j$, namely $\alpha_i$ and $\alpha_j$, respectively and $a$ is the damping factor.
\\
In contrast to organic-inorganic lead halide perovskite materials such as $\mathrm{MAPbI}_{3}$ and $\mathrm{FAPbI}_{3}$, in the case of $\mathrm{CsPbI}_3$, that is purely constituted of elemental ions, there are no permanent atomic dipoles and quadrupoles.  
The induction part however still plays a role in $\mathrm{CsPbI}_3$ as a consequence of structural anisotropies and finite temperature distortions. In the AMOEBA FF, the induced dipole $\boldsymbol{\mu}^{\mathrm{ind}}$ is evaluated self-consistently as follows:
\begin{equation}\label{induce_dipol}
\mu_{i,\beta}^{\mathrm{ind}}=\alpha_i\Bigg(\sum_{\{j\}}T_{ij,\beta}\mathbf{M}_{j}+\sum_{\{j'\}}T_{ij',\beta\gamma}\mu_{j',\gamma}^{\mathrm{ind}}\Bigg) \quad\quad\mathrm{for} \quad\quad \beta,\gamma=x,y,z.
\end{equation}
Here the set of atomic coordinates $\{j\}$ includes all the atomic sites outside the molecule which contains $i$, whereas the set of atomic coordinates $\{j^{'}\}$ contains all the atomic sites except $i$ itself. In the case of CsPbI$_3$ the induced dipole equation simplifies to $\mu_{i,\beta}^{\mathrm{ind}}=\alpha_i[\sum_{\{j\}}T_{ij,\beta}q_{ij}+\sum_{\{j'\}}T_{ij,\beta\gamma}\mu_{j,\gamma}^{\mathrm{ind}}]$.\\
For the van der Waals interaction term $U^{\mathrm{vdW}}$, we use a  Lennard-Jones (LJ) potential with parameters $\epsilon_{ij}$ and $\sigma_{ij}$, as opposed to other possible choices such as Buckingham\cite{buckingham_classical_1938} or Hill potentials\cite{hill_steric_1948}. These three interatomic potentials differ in their repulsive parts, which could have an effect on e.g. the interaction between nearest I-Pb ions that lie in a typical range of 3.1-3.5 $\AA$ in lead halide perovskite materials \cite{deretzis_exploring_2017}. However, it turns out that the functional form of the LJ potential appears to have sufficient flexibility to achieve close matches between the FF forces and the first-principles reference data.

Taken all together for the case of CsPbI$_3$, the {\it pol} FF, has the following form:
\begin{equation}\label{FFpolar}
U^{\mathrm{pol}}(r_{ij})=4\varepsilon_{ij}\Bigg[\bigg(\frac{\sigma_{ij}}{r_{ij}}\bigg)^{12}-\bigg(\frac{\sigma_{ij}}{r_{ij}}\bigg)^{6}\Bigg]+\frac{1}{4\pi\varepsilon_0}\Bigg[\frac{q_iq_j}{r_{ij}}- q_i \frac{\mathbf{r}_{ij}\cdot\boldsymbol{\mu}_{j}^{\mathrm{ind}}}{r_{ij}^3} - q_j \frac{\mathbf{r}_{ij}\cdot\boldsymbol{\mu}_{i}^{\mathrm{ind}}}{r_{ij}^3} +(\boldsymbol{\mu}_i^{\mathrm{ind}})^{\mathrm{T}}\mathbf{T'}_{ij}(\boldsymbol{\mu}_j^{\mathrm{ind}})\Bigg].
\end{equation}
Here $\mu_{ij}^{\mathrm{ind}}=[\mu_{ij,x}, \mu_{ij,x}, \mu_{ij,x},]^{T}$ and $\mathbf{T'}_{ij}$ is a $3\times3$ matrix composed by elements with form $T_{ij,\beta\gamma}$.   
For the {\it npol} variant,
we use the same functional form given in Eq.~\ref{FFpolar} but omit all the terms involving $\boldsymbol{\mu}^{\mathrm{ind}}$.\\

\subsection{Fitting procedure}
To determine the parameters of the {\it npol} and {\it pol} FFs for $\mathrm{CsPbI}_3$, a FM approach has been employed. Originally introduced by  Ercolessi and  Adams in 1994 \cite{ercolessi_interatomic_1994}, this method is a powerful approach to directly construct interatomic potentials from first-principles calculations in a semi-automated way. The central idea is the minimization of an objective function, which in the current work is defined as follows:
\begin{equation}\label{forcematching}
\tau_{\text{F}}\left(\{\boldsymbol{\sigma}\}\right)=\bigg(\sum_{k=1}^{\text{M}} \text{N}_{k}\bigg)^{-1} \sum_{k=1}^{\text{M}}\sum_{i=1}^{\text{N}_{k}}\Big| \mathbf{F}_{ki}(\{\boldsymbol{\sigma}\})-\mathbf{F}_{ki}^0\Big|^2. 
\end{equation}
In this non-linear least squares problem, the goal is to determine the set of parameters $\{\boldsymbol{\sigma}\}$ which minimize the difference between the forces from first-principles $\mathbf{F}_{ki}^0$ and the forces $\mathbf{F}_{ki}(\{\boldsymbol{\sigma}\})$ that are constructed from the interatomic potential employing the set of parameters $\{\boldsymbol{\sigma}\}$. The first-principles reference forces are generated from DFT molecular dynamics runs at different temperature and pressure conditions. Therefore, in Eq. \ref{forcematching}, the index $k=1,...,\mathrm{M}$ denotes the trajectory frame under consideration with M being the total number of frames, and the index $i=1,...,\mathrm{N}_k$ represents the atom in frame $k$ where $\mathrm{N}_k$ is the total number of atoms (in frame $k$) included in the fitting procedure. Therefore $\big(\sum_{k=1}^{\text{M}} \text{N}_{k}\big)^{-1}$ is the normalization factor, i.e. the inverse of the total number of atomic forces for all atoms and frames that are used for the fitting.

Other physical quantities, like e.g. energies and bulk moduli, can also be included in the objective function in addition to the forces themselves.\cite{ercolessi_interatomic_1994, li_embedded-atom-method_2003} However here, we opted for a purely force-based objective function as given in Eq. \ref{forcematching}.

For the practical implementation of the FM approach, an interface between a classical molecular dynamics and a minimization package has to be developed.  For the latter, the minimization package  \texttt{MINPACK} \cite{more_user_1980,more_minpack_1984} was selected, which offers numerically stable subroutines to solve linear and non-linear least squares problems, while for the former the classical molecular dynamics software TINKER \cite{Ponder_Tinker_2010,rackers_tinker_2018} was adopted, which was originally designed for the polarizable AMOEBA FF but enables also the use of a variety of other FFs. In this way, at each iteration $I$, the classical forces $\mathbf{F}_{ki}(\{\boldsymbol{\sigma}^{I}\})$ are computed with a given set of parameters $\{\boldsymbol{\sigma}^I\}$ by TINKER and then used to compute the modulus squared of the force difference $| \mathbf{F}_{ki}(\{\boldsymbol{\sigma}^I\})-\mathbf{F}_{ki}^0|^2$ as a contribution to the objective function in Eq.~\ref{forcematching} that at the end of the iteration is minimized with a Levenberg-Marquardt algorithm from MINPACK to determine a new set of parameters $\{\boldsymbol{\sigma}^{I+1}\}$. This iterative process is pursued until the convergence parameters of MINPACK lie below $10^{-9}$.

\subsection{Computational details }

\subsection{Static DFT Calculations} \label{system_optimiz}

To determine the supercell size and to optimize the cell and geometry of the $\delta$ and $\alpha$ phase of CsPbI$_3$, we employed the Quantum ESPRESSO (QE) package version 6.0 \cite{giannozzi_quantum_2009}. 
DFT calculations were performed with the exchange-correlation functional by Perdew, Burke and Ernzerhof (PBE) \cite{perdew_generalized_1996}. For Pb and I, we used pseudopotentials of the Rappe-Rabe-Kaxiras-Joannopoulos type \cite{rappe_optimized_1990}, while, for Cs, which has only one valence electron, we used an ultrasoft pseudopotential of Vanderbilt type \cite{vanderbilt_soft_1990}. The plane wave cutoff selected for the expansion of the wavefunction and charge density are 50 Ry and 400 Ry, respectively, each determined employing a convergence criterion of $\sim 10^{-3}$ Ry per stoichiometric unit (s.u.) for the total energy. To determine the supercell sizes we made use of k-point sampling and adopted convergence criteria of $\sim 10^{-3}$ Ry on the total energy per stoichiometric unit (s.u.) and $\sim 10^{-4}$ Ry/a.u. for the atomic forces. Cell and geometry optimizations have been performed on supercells of sizes 2x4x1 for $\delta$ CsPbI$_3$ (160 atoms) and 4x4x3 for $\alpha$ CsPbI$_3$ (240 atoms) constructed from the experimental unit cell of Ref.~\cite{stoumpos_semiconducting_2013} and Ref.\cite{trots_high-temperature_2008}, respectively. In the remainder of the manuscript, the cell volumes obtained through optimizations are referred to as $V_0$. In order to include higher flexibility in the FM procedure, we also considered additional supercells with a different volume, which we refer to as $V_1$. The latter cells were constructed by isotropical expansion of the 0K optimized $\delta$ and $\alpha$ phase lattices by 5\% on each axis. The cell volume and lattice parameters are summarized for the two phases in Table~\ref{Volume_optimized}.

\subsubsection{Generation of first-principles MD trajectories}

Using Car-Parinello (CP) molecular dynamics within the CPMD software version 4.1 \cite{andreoni_new_2000}, trajectories have been generated in the NVE ensemble for $\alpha$ CsPbI$_3$ at an average temperature of 650 K and for $\delta$ CsPbI$_3$ at 100, 300 and 500 K, for both volumes V$_0$ and V$_1$ 
using the PBE exchange-correlation functional and pseudopotentials of Goedecker-Teter-Hutter type \cite{goedecker_separable_1996,hartwigsen_relativistic_1998,krack_pseudopotentials_2005}. A cutoff of 90 Ry was used for the plane-wave expansion of the Kohn-Sham orbitals determined by applying a convergence criterion of $\sim 10^{-3}$ Ry per s.u. for the total energy. Before starting the MD, we performed an additional geometry optimization with the CPMD software. The time step adopted for the CP dynamics was 5 a.u. and the fictitious electron mass was set to 800 a.u. The systems were first equilibrated at the different temperatures applying a Nos\'e-Hoover thermostat \cite{nose_molecular_1984,hoover_canonical_1985} with a coupling frequency of 1000 cm$^{-1}$ for $\sim 1$ ps followed by NVE runs of the equilibrated systems of $\sim 7$ ps. From these trajectories, we collected the frames to perform the FM. 

\begin{table}[h!]
\centering
\caption{Optimized (0K) cell dimensions of the $\alpha$ and $\delta$ phases compared with the experimental unit cell. In CP trajectories and NVT classical trajectories, the same cell is used at finite temperature, while for NPT trajectories information about the volume variation at finite temperature is given in the Supporting Information. Cell dimensions are given in terms of the number of unit cells replicated in each direction, and $\Delta$V \% stands for the relative difference of the volume per s.u. with respect to the experimental one. RMSD is the root mean square deviation from the experimental structure. }\label{Volume_optimized}
\begin{tabular}{lccccc}
\hline
\multicolumn{6}{c}{$\alpha$ phase} \\
\hline
    & Experimental\cite{trots_high-temperature_2008} & DFT-V$_0$  & DFT-V$_1$ & $npol$ & $pol$ \\
\hline
(super)cell size        & 1x1x1  & 4x4x3         & 4x4x3  & 7x7x7        & 7x7x7       \\
a=b [\AA]               & 6.29   & 25.43 (6.36)  & 26.70  & 43.63 (6.23) & 43.88 (6.27)\\
c  [\AA]                &  =a=b  & 19.36 (6.45)  & 20.33  & =a=b         & =a=b        \\
V per s.u. [\AA$^3$]    & 248.86 & 260.83        & 301.94 & 242.14      & 246.32      \\
$\Delta$V \%            & -      & +5\%          & +21\%  & -3\%        & -1\%.       \\
RMSD [\AA]              & -      & 0.45          & -      & 0.20        & 0.08        \\
\hline
\multicolumn{6}{c}{$\delta$ phase} \\
\hline
    & Experimental \cite{stoumpos_semiconducting_2013} & DFT-V$_0$  & DFT-V$_1$ & $npol$ & $pol$ \\
\hline
(super) cell size       & 1x1x1  & 2x4x1          & 2x4x1  & 4x8x2         & 4x8x2        \\
a [\AA]                 & 10.43  & 21.60 (10.80)  & 22.68  & 40.55 (10.14) & 42.78 (10.70)\\
b  [\AA]                & 4.79   & 19.56 (4.89)   & 20.54  & 40.12 (5.02)  & 38.90 (4.86) \\
c  [\AA]                & 17.76  & 18.25          & 19.16  & 36.43 (18.22) & 35.21 (17.61)\\
V per s.u. [\AA$^3$]    & 221.82 & 240.95         & 278.93 & 231.51        & 228.88       \\
$\Delta$V \%            & -      & +9\%           & +26\%  & +4\%          & +3\%          \\
RMSD [\AA]              & -      & 0.29           & -      & 0.78          & 0.39          \\
\hline
\end{tabular}
\end{table}

\subsubsection*{Force matching}
To perform the FM, we used the interface between the MINPACK-1 package\cite{more_user_1980,more_minpack_1984} and TINKER\cite{rackers_tinker_2018} described in the section about the fitting procedure.
The input forces are 
collected from a total of 60 frames: 12 frames for each of the two $\alpha$ phase trajectories each containing 240 atoms, and 6 frames for each of the six $\delta$ phase trajectories with 160 atoms each. With this choice, 2880 forces are given as input to the interface for the $\alpha$ phase and the same number for the $\delta$ phase. The frames have been sampled equidistantly with an interval of $0.48$ ps. To verify that the number of frames was sufficient, we also performed the FM for a smaller and a larger number of frames, and no substantial differences were observed. To obtain accurate forces as input for the FM interface, we performed for each frame an additional self-consistent DFT cycle employing a larger plane wave cutoff for the wave function expansion of 110 Ry. In addition to the forces, initial guesses of the FF parameters have to be given as an input. For this, we started using parameters of existing FFs such as the two sets of parameters developed by Mattoni and coworkers for MAPbI$_3$\cite{mattoni_methylammonium_2015,caddeo_collective_2017}, re-scaled parameters coming from MgSiO$_3$\cite{matsui_computational_1987,pinilla_atomistic_2017}, and parameters given by single ions interacting with solvent %\cite{jain_electronic_1976,}
\cite{joung_determination_2008,li_rational_2013}.
The fitting procedure was started from several different sets of initial parameters and the chosen final result was the one that led to i) the minimal root-mean-square deviation of the force magnitude \cite{fellinger_force-matched_2010}\
\begin{equation}\label{error_forces}
    |\Delta \mathbf{F}|_{\mathrm{rms}}=\sqrt{\sum_{k=1}^{M}\sum_{i=1}^{N_k}\omega_{ki}\Bigg(\frac{|\mathbf{F}_{ki}^{\mathrm{FF}}|-|\mathbf{F}_{ki}^{0}|}{|\mathbf{F}_{ki}^0|}\Bigg)^2}, % \times 100 \%,
\end{equation}
where $|\mathbf{F}_{ki}^{\mathrm{FF}}|$ is the magnitude of the forces computed with the FF model and $\omega_{ki}$ the weight that depends on the magnitude of the DFT forces:
\begin{equation}\label{error_angle}
    \omega_{ki}=\frac{|\mathbf{F}_{ki}^0|}{\sum_{k'=1}^{M}\sum_{i'=1}^{N_{k'}}|\mathbf{F}_{k'i'}^0|}.
\end{equation}
 and that ii) Minimized the angular deviation of the force directions
 \begin{equation}
     \theta_{\mathrm{avg}}=\sum_{k=1}^{M}\sum_{i=1}^{N_{k}}\omega_{ki}\theta_{ki},
 \end{equation}
 where $\theta_{ki}$ is the angle between the DFT forces and the forces computed with the FF model. As a first test of the resulting FF, the relative energetic ordering at 0 K of the $\delta$ phase with respect to the $\alpha$ phase was determined. For this, the total energy for each phase was computed with TINKER after optimizing the cell and geometry with a given set of FF parameters. More details about this procedure are given in the next section. 
 
 Overall we observed that the vdW parameters resulting from the FM procedure are very sensitive to the choice of the input parameters, while the electrostatic parameters are more stable. 
 
 With this fitting procedure, we developed a polarizable FF for CsPbI$_3$ with the form given in Equation~\ref{FFpolar} ({\it pol} model) and a fixed-point charge ({\it npol} model) FF, which includes only the permanent electrostatic potential. In total, we determined 16 parameters for the {\it pol} FF and 13 for the {\it npol} FF. Since, the effect of the damping factor $a$ is likely to be of minor importance in a crystal like CsPbI$_3$, we did not include this parameter in the fit. Instead, we used its default value of 0.39. \cite{ren_polarizable_2011}
 Charge neutrality was imposed during the fit by setting $q_{\mathrm{Cs}}=q_{\mathrm{Pb}}+q_{\mathrm{I}}$.
 The vdW parameters $\sigma$ and $\epsilon$ have been determined explicitly for the Pb-I and Cs-I pair interactions, while for Cs-Pb, they are computed internally by TINKER via the Lorentz-Berthelot mixing rules:
\begin{equation}\sigma_{\mathrm{PbCs}}=\frac{\sigma_{\mathrm{Pb}}+\sigma{\mathrm{Cs}}}{2},\;\; \varepsilon_{\mathrm{PbCs}}=\sqrt{\varepsilon_{\mathrm{Pb}}\varepsilon_{\mathrm{Cs}}}.
\end{equation} 
This is justified by the larger interatomic distance between Cs-Pb, which leads to relatively small vdW contributions without a major impact on the fitting results.

\subsubsection{FF based MD runs}
All classical MD simulations were performed with the TINKER code.\cite{rackers_tinker_2018} For the {\it npol} model, we used a vdW cutoff ($R_{\textrm{vdW}}$) of 17 \AA\  and a real-space cutoff ($R_{\textrm{EW}}$) for the Ewald summation of the electrostatic interactions of 6 \AA.
For the {\it pol} model, $R_{\textrm{EW}}$ was the same while $R_{\textrm{vdW}}$=16 \AA.
The system size was chosen in such a way as to respect the minimum image convention, i.e. the edge of the supercell was at least double the largest cutoff used in the simulation. The simulation systems thus contained  4x8x2 (1280 atoms) for the $\delta$ phase and 7x7x7 (1715 atoms) for the $\alpha$ phase.
Before performing MD simulations, we optimized the cell and the geometry and used the total energy of the fully optimized systems to determine the energetic ordering of the two phases at 0 K. For the $\alpha$ phase we also obtained an optimized structure starting from frames generated from the NPT trajectories at 100K. We performed the cell and coordinate optimizations for frames at 0.1 ns intervals and we chose the one lowest in energy. The use of so few frames is justified by the small order of magnitude of energy difference (below 0.02 kcal/mol per s.u.) between the obtained optimized structures. To assess the performance of the developed {\it pol} and {\it npol} FFs at finite temperature, we generated trajectories for the $\alpha$ and $\delta$ phase at 100 K, 300 K, 500 K, 600 K and 650 K, in the NVT and NPT (with P=1 atm) ensembles. For temperature control, the Bussi thermostat with a coupling time of 0.1 ps and a Monte Carlo barostat with a coupling time of 2 ps were used. The employed time step was 2 fs and after equilibration,  trajectories were run for 1 ns.

\subsubsection{Evaluation of system properties}
{\it Comparison with Bader charges}

For each ab initio trajectory, the coordinates of twelve snapshots were taken at equidistant time intervals of 0.48 ps and Gaussian CUBE files were generated from the optimized density using the \textit{PostProc} package from QE version 6.6.\cite{giannozzi_quantum_2020,giannozzi_quantum_2009,giannozzi_advanced_2017} The density used for generating the cube file was computed using the PBE functional\cite{perdew_generalized_1996} with a wavefunction cutoff of 90 Ry and a charge density cutoff of 720 Ry as well as employing a 3x3x3 Monkhorst k-point grid. The pseudopotentials are the same as described in Sec~\ref{system_optimiz}. From the obtained cube files Bader charges were computed using the program developed by Henkelman et al.\cite{henkelman_fast_2006}
\\
\\{\it Analysis of structural and energetic properties}

We analyzed the radial pair distribution functions (RPDF) for all the different pairs of atoms, for both the optimized supercells at 0K and the trajectories at finite temperature. We compute the RPDF with VMD~\cite{humphrey_vmd_1996}, considering spherical slices of bin size 0.1 \AA\ at finite temperature and 0.01 \AA\ at 0K. For the DFT trajectories, we considered equidistant frames taken every 1.2$\times 10^{-2}$ ps and for the classical trajectories every 1 ps. To analyze the energies and the volumes of the classical trajectories at finite temperature, we computed the time average from frames taken every 1 ps.\\ 
\\
{\it Analysis of vibrational properties}

To compute the CsPbI$_3$ power spectra we made use of the TRAVIS code \cite{brehm_travis_2011}. For the DFT trajectories, we analyzed 550 frames taken every 1.2$\times 10^{-2}$ ps while for the classical MD we used a time interval of 0.1 ps. In both cases, the autocorrelation function has been weighted with respect to the atomic masses. For a better comparison of the shape of the power spectra at different temperatures, the spectra have been normalized to 1. Additionally, we smoothened the power spectra generated by the Travis code for the classical trajectories employing the Gaussian filter as implemented in Scipy with \texttt{sigma}=20.

\section{Results and discussion}

\subsection{Force-matched force fields }

The parameters obtained from the FM procedure for the {\it pol} (Eq.~\ref{FFpolar}) and {\it npol} FF are summarized in Table \ref{tbl:param}. 
\begin{table}
  \caption{Parameters of the {\it pol} and {\it npol} FFs for CsPbI$_3$ obtained via FM. }
  \label{FF_parameters}

  \begin{tabular}{lcccc|ccc}

    \hline
    &\multicolumn{4}{c|}{{\it pol}} & \multicolumn{3}{c}{{\it npol}} \\
    \hline
    & $\sigma$ [\AA] & $\varepsilon$ [kcal/mol]\ &$q$ [e]& $\alpha$ [\AA$^3$] & $\sigma$ [\AA] & $\varepsilon$ [kcal/mol]& $q$  [e] \\
    \hline
    Pb & 5.1647 &0.1140& 1.1860&  2.720& 4.8620& 0.1271  &  0.8589  \\
    Cs & 4.2631& 0.3295& 0.7919 &0 & 3.4910 & 3.4422& 0.7680   \\
    I &4.6452 & 0.0544 &  -0.6593& 5.050 & 4.6121& 0.0611 & -0.5423 \\
    Pb-I &3.9357 &0.0620&- &- & 3.8623 & 0.0499& -   \\
    Cs-I &4.2053 & 0.1448&- &- & 4.2024& 0.1403&-    \\
    \hline
  \end{tabular}\label{tbl:param}
\end{table}
At difference to its {\it npol} counterpart, the  {\it pol} model also includes the interactions between atomic charges and induced dipoles as well as the interactions among induced dipoles. As shown in Eq.~\ref{induce_dipol}, the induced dipole is proportional to the polarizability $\alpha$, which is the only additional parameter of the {\it pol} FF. For both FFs,  the resulting effective atomic charges are quite far from the values for the fully isolated ions ($q_{\mathrm{Pb}}$=+2 e, $q_{\mathrm{Cs}}$=+1 e and $q_{\mathrm{I}}$=-1) but show a good agreement with average Bader charges computed from the electron density distributions along the DFT trajectories ($q_{\mathrm{Pb}}$=+0.936 $\pm$ 0.018 e, $q_{\mathrm{Cs}}$=+0.839 $\pm$ 0.015 e and $q_{\mathrm{I}}$=-0.592 $\pm$ 0.007 e). In fact, for Pb and I, the calculated Bader charges are in between the effective charges of the {\it pol} and {\it npol} models while the one of Cs is slightly larger in both FFs, but the difference between the three values is within 0.1 e. Moreover, the differences of the averaged Bader charges obtained for the individual phases, temperatures and cell sizes, are as small as  0.048 e for all three ions (values in the Supporting Information), which is a first indication that, at least from an electrostatic point of view, a single FF model should indeed be able to simultaneously reproduce the different phases of CsPbI$_3$ at finite temperature.

The experimental values of the polarizabilities are 2.44 \AA$^3$ for Cs$^+$, 7.16 \AA$^3$ for I$^-$ \cite{solomonik_effective_1979} and 2.02 \AA$^3$ for Pb$^{2+}$ \cite{hanni_polarizabilities_2010}. Our parameters do reproduce the physical trend between Pb$^{2+}$ and I$^-$, although, for the latter, the atomic polarizability is around 2 \AA$^3$ smaller. The largest difference exists for Cs$^+$, which in the FF case has zero polarizability. This might be due to the fact that the Cs ions are free to move within the Pb-I cages, and the induced dipole moment might be partially cancelled out by their thermal motion.

The vdW interactions were fitted using 10 parameters for both FFs (2 for each of the 3 species and another 4 for the explicitly treated Pb-I and Cs-I pair interactions). As already mentioned, the fitting procedure is less sensitive to the values of the (weaker) van der Waals interactions and the resulting $\sigma$ and $\varepsilon$ values follow less of a physical trend than the effective charges. 

Overall, in the {\it npol} case, the objective function in Eq.~\ref{forcematching} could be minimized down to a relative average error between first-principles and FF forces of $|\Delta F|_{\mathrm{rms}}=39.5\%$ and an angular deviation of the force directions of $\theta_{\mathrm{avg}}=26.8^\circ$, whereas for the {\it pol} model, the deviations are further reduced to $|\Delta F|_{\mathrm{rms}}=32.7\%$ and $\theta_{\mathrm{avg}}=16.6^\circ$. These errors lie in a range typically observed for the FM procedure ~\cite{li_embedded-atom-method_2003, fellinger_force-matched_2010}. 
Comparing the two FFs, the {\it pol} model clearly shows a lower error than {\it npol}. Particularly interesting is the decrease of $10.2^\circ$ in the directional force error $\theta_{\mathrm{avg}}$, which might be a reflection of the fact that the {\it pol} FF accounts for anisotropic charge distributions.

\subsection{Zero Kelvin Properties}
\subsubsection{Structural properties at 0K} \label{0Kstructuralporperties}

\begin{figure}[h!]
    \centering
    \includegraphics[width=7cm]{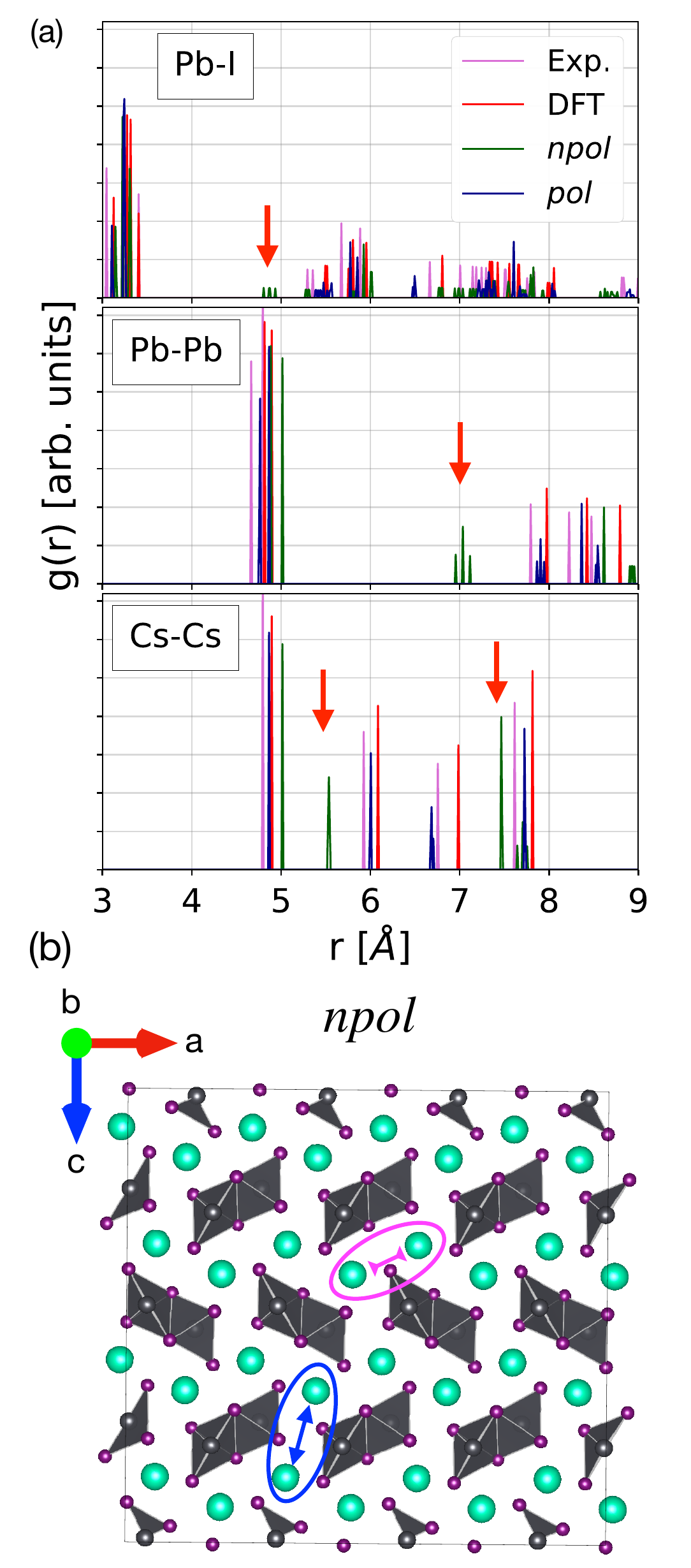}
    \caption{Structural properties of the optimized $\delta$ phase: (a) RPDFs of the $\delta$ phase optimized using DFT and the {\it npol} and {\it pol} FFs compared to the experimental RPDFs for the pairs Pb-I, Pb-Pb and Cs-Cs. The red arrows indicate peaks obtained with the {\it npol} model with the largest deviations in comparison to those from other levels of theory and experiments. (b) {\it npol} atomistic structure. The red arrow shows the deviations in the separation of face-sharing octahedral pillars. The Cs-pairs highlighted in magenta get closer in the {\it npol} structure while the pair indicated in blue are separated further apart.}
    
    \label{Figure2}
\end{figure}

The FF parameters obtained with the FM method (Table \ref{FF_parameters}) have been used to optimize the cell and the geometry of the $\alpha$ and $\delta$ phases and the results are given in Table \ref{Volume_optimized}. To assess the predictive performance of the FF in describing the $\alpha$ and $\delta$ structure, we use the experimental structure as the reference (Table ~\ref{Volume_optimized}). However, since the FF parameters were fitted based on DFT data, the theoretical performance of the FF should also be judged with respect to the DFT data.  
For the $\alpha$ phase, the FFs generate 0K optimized cells with slightly smaller volumes per s.u. than the one of the experimental unit cell (measured at 634 K) with deviations of -3\% and -1\% for the {\it npol} and the {\it pol} models, respectively. The volume of the $\delta$ phase measured at room temperature is slightly larger, showing an increase of +4\% with the {\it npol} model and +3\% with the {\it pol} model. In both cases, the obtained deviations are smaller than the ones of the DFT optimizations which are +5\% for the $\alpha$ phase and +9\% for the $\delta$ phase. Although both FFs seem to reproduce the overall volume of the experimental $\alpha$ and $\delta$ phase in a comparable manner, if one takes into account the individual unit cell axes (Table \ref{Volume_optimized}, results in parenthesis), one can observe that the contraction of the $\alpha$ phase volume is isotropic, in contrast with the $\delta$ phase, in which some of the axes expand and others contract. The opposite is observed with the DFT method, in which in the $\alpha$ phase the axes become inequivalent, most probably as an artifact of the anisotropic shape of the supercell chosen to reduce computational cost.
The root mean square displacement (RMSD) between the experimental and the optimized structures averaged over all the supercells (Table \ref{Volume_optimized}), computed with the VMD software \cite{humphrey_vmd_1996}, shows
for the $\alpha$ phase the same trend as for the volume deviation, with the DFT geometry having the worst agreement. In contrast, for the $\delta$ phase the trends do not coincide, \text{e.g.}, the DFT geometry has the smallest RMSD but the largest volume expansion. 
The worst agreement is obtained for the $\delta$ phase structure optimized with the {\it npol} model. An idea about the atomistic origin of this disagreement can be obtained from an analysis of the radial pair distribution functions (RPDFs). While most of the RPDFs are in close agreement (Figure S1 in the Supporting Information), some larger peak displacements among the different models are observed for Pb-I, Pb-Pb and Pb-Cs pairs in the $\delta$ phase (Figure \ref{Figure2}). In fact, from the RPDFs of Pb-I shown in Figure \ref{Figure2}a, it is evident that all models agree relatively closely with respect to the position of the first peak (corresponding to the length of the Pb-I bonds) but there are clear differences in the location of the second-nearest neighbor interactions, where the {\it npol} model has three small peaks at $\sim$4.9\AA, which is displaced by as much as $\sim$0.5\AA~from the position of the second coordination sphere of the other models. A similar discrepancy is observed for the second shell of the Pb-Pb RPDFs, where the peaks usually situated around $\sim$8\AA~are displaced by $\sim$1\AA~to $\sim$7\AA~in the {\it npol} model, and in the Cs-Cs RPDFs where the second-coordination shell peak that should be located at $\sim$6\AA\; is at $\sim$5.5\AA\ instead, whereas the peak that should be centered at $\sim$6.8\AA~ is displaced in the opposite direction by $\sim$0.7\AA~ (to $\sim$7.5\AA). These differences in the RPDFs are the consequences of changes in the relative distances between edge-sharing octahedral pillars of the $\delta$ phase and the Cs ion displacements in the interoctahedral spaces as visualized in Figure \ref{Figure2}b. This figure shows that pairs of Cs ions such as the two highlighted in magenta get closer together (by 0.33 \AA~ with respect to the experimental reference) while those equivalent to the one in blue are further separated (by 0.80 \AA). Nevertheless, even though the structure is not reproduced in all fine details, it retains the overall characteristics of the CsPbI$_3$ $\delta$ phase, in which the octahedra are edge-sharing.  In the next section, we will show that the deviation of the optimized $\delta$ phase structure obtained with the {\it npol} FF also leads to some discrepancies in the finite temperature properties. 

In contrast, regarding the $\alpha$ phase, the RPDFs of all models are in good agreement as shown in Figure S2 in the Supporting material.

\subsubsection{Energy analysis at 0K}
The newly-developed FFs reproduce the correct energetic ordering of the two phases when optimized starting from experimental structures~\cite{trots_high-temperature_2008,stoumpos_semiconducting_2013} with the $\delta$ phase being more stable than the $\alpha$ phase. The energy difference between the $\delta$ and $\alpha$ phases at the \textit{ab initio} level is taken as reference with $\Delta E=E_{\alpha}-E_{\delta}=3.90$ kcal/mol per s.u.. The polarizable model reproduces the energy difference in excellent agreement ($\Delta$E = 3.99 kcal/mol per s.u.) while the {\it npol} model yields the correct energetic order but clearly underestimates the energy difference ($\Delta E$=0.36 kcal/mol per s.u.). From an analysis of the total energy as a function of temperature (see Sec.~\ref{finite_T_energy}) (that in general follows a roughly linear trend), we realized that the FF-optimized $\alpha$ phase structure obtained from the experimental data likely corresponds to a local minimum. In fact, by optimizing the structure starting from frames of the NPT trajectories at 100 K, due to symmetry breaking of the perfectly cubic experimental structure, we obtain a structure that is lower in energy and that presents an axes ratio and octahedral tilting reminiscent of the orthorhombic phase (Supporting Information).  
In the {\it npol} FF, due to the small energy difference between the two phases, this new orthorhombic structure leads to a reversal of the energetic ordering between $\alpha$ and $\delta$ phases. In contrast, in the {\it pol} model the energy difference is reduced by 1.57 kcal/mol per s.u. but the ordering is maintained. This reduction should not be interpreted as a disagreement with respect to the DFT reference, which was itself optimized starting from the highly-symmetric cubic experimental structure.

Details of the energetics of the two phases, including the $\alpha$ phase structures optimized from 100K trajectories frames, are given in Table \ref{Energy}, in which the values and the differences of the total energy and its components, such as vdW ($E_{\mathrm{vdW}}$), permanent electrostatic components ($E_{\mathrm{el,perm}}$) and the induced electrostatic contribution ($E_{\mathrm{el,ind}}$) of the polarizable model are shown.  

Not surprisingly, the contributions from electrostatic interactions are dominant but due to the small energy difference between the phases, the vdW interactions can become of comparable influence and can have a decisive effect in some cases.
In fact, in both models, the vdW contribution is destabilizing the $\alpha$ phase and favoring the $\delta$ phase, thus counteracting the permanent electrostatic contribution. In the {\it npol} model, vdW and electrostatic energy difference contributions are of similar size (though with opposite signs). In the FF re-optimized experimental structures, the former is slightly larger leading to a more stable $\delta$ phase, while for the new minimum of the $\alpha$ structure derived from the 100 K dynamics, the opposite is the case. In the {\it pol} model on the other hand, the vdW energy difference is very small, even smaller in the 100 K derived structure, and it is primarily the difference in $E_{\mathrm{el,ind}}$ that is fundamental to obtain the correct energetic order. The reason for this is that in the highly symmetric $\alpha$ phase, the induced electrostatic energy is negligible or very small (100 K structure) while in the  anisotropic $\delta$ phase, the induced dipoles and thus  $E_{\mathrm{el,ind}}$ are substantial. This clearly demonstrates that the inclusion of effects due to induced electrostatic contributions, such as in the {\it pol} FF, is fundamental for a correct description of the energetic ordering of the two phases. 
      
\begin{table}[]
\centering
\caption{Total energy and its contributions (in kcal/mol per s.u.) of the $\alpha$ and $\delta$ phase optimized with the {\it npol} and {\it pol} models at 0K. For the $\alpha$ phase the column labeled by (exp.) corresponds to the re-optimized experimental structure, while the one labeled (100 K) refers to the structure obtained from the optimization of the frames of the 100 K trajectory. The energy difference is computed as $\Delta E=E_{\alpha}-E_{\delta}.$}\label{Energy}
\begin{tabular}{lccccc}
\hline
    & \multicolumn{5}{c}{{\it npol}}  \\
    & $\alpha$ phase (exp.) & $\alpha$ phase (100 K) & $\delta$ phase & $\Delta$E (exp.)&  $\Delta$E (100K) \\
\hline
$E_\mathrm{tot}$  & -172.90%    
& -174.17  & -173.26 %-173.28      
& 0.36%8
& -0.91\\
$E_\mathrm{vdW}$  & 3.78      & 3.29   & 2.38%4         
& 1.40%4   
& 0.91\\
$E_\mathrm{el,perm}$  & -176.68   & -177.46  & -175.64%2
& -1.04%6  
& -1.82\\
\hline
    &  \multicolumn{5}{c}{{\it pol}}  \\
    & $\alpha$ phase (exp.) & $\alpha$ phase (100 K) &  $\delta$ phase & $\Delta$E (exp.)&  $\Delta$E (100K) \\
\hline
$E_\mathrm{tot}$  & -258.10  & -259.67 & -262.09      & 3.99  & 2.42  \\
$E_\mathrm{vdW}$  & 12.62     & 12.02   & 12.01        & 0.61  & 0.01  \\
$E_\mathrm{el,perm}$ &  -270.72  & -271.05    & -265.79      & -4.925  & -5.26 \\
$E_\mathrm{el,ind}$ & 0.00   & -0.64     & -8.30        & 8.30  & 7.66\\
\hline
\end{tabular}
\end{table}

\subsection{Finite temperature properties}

\subsubsection{Radial pair distribution functions}
For NVT simulations we obtain (meta)stable phases at all investigated temperatures, while in the NPT simulations, only the {\it npol} model is able to generate $\alpha$ phase trajectories at 600 K and 650 K  which do not melt (the experimental melting temperature is expected to be above 700 K~\cite{trots_high-temperature_2008}).

To determine the ability of the {\it npol} and {\it pol} FFs in reproducing the structural properties of the DFT reference at finite temperature, we analyze the RPDFs of all systems for which the structures remained intact. Since, in the case of the  classical MD simulations, the NPT trajectories are the most realistic ones, we will focus on these trajectories, the corresponding RPDFs of the NVT simulations are instead included in the Supporting Information (Figures S3-S8). Overall the differences between the two are minor.

\begin{figure}
    \centering
    \includegraphics[width=18cm]{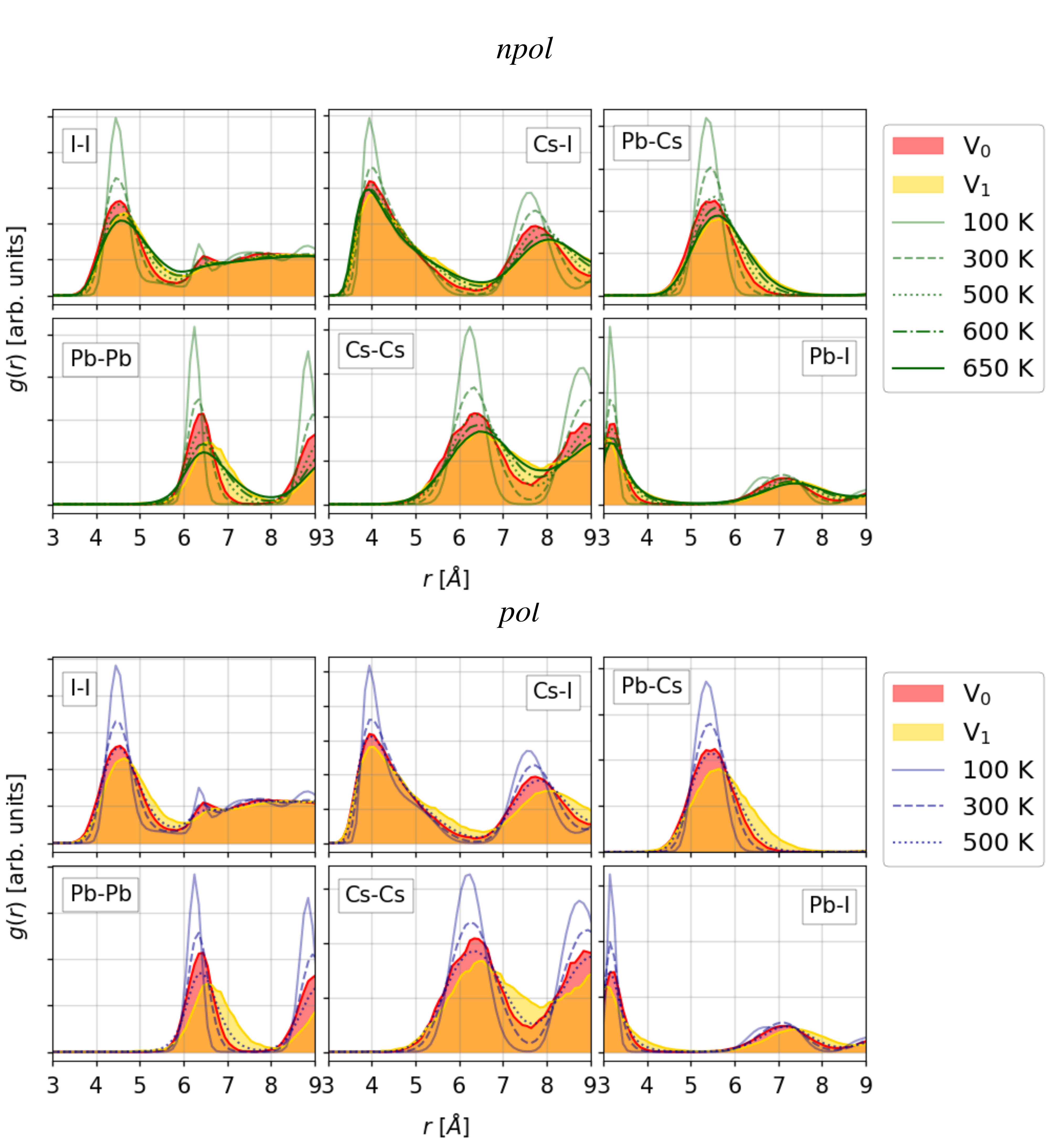}
    \caption{RPDFs of the $\alpha$ phase trajectories generated at different temperatures in the NPT ensemble. In the top panel, the RPDFs are shown for the {\it npol}  model up to T = 650K. In the bottom panel, the RPDFs are shown for the {\it pol}  model up to T = 500K, since for higher temperatures the system is melting. Note that the DFT RPDFs for the volume $V_0$ and $V_1$, are computed from trajectories generated at 650K.}\label{cubic_NPT}
    
\end{figure}

In section~\ref{0Kstructuralporperties} dedicated to the structural properties at 0 K, we have noted that both the {\it npol} and {\it pol} models can closely reproduce the RPDF at 0 K obtained from DFT data as well as the experimentally determined structure of the $\alpha$ phase. The same is observed at finite temperatures. In Figure~\ref{cubic_NPT}, the RPDFs of the NPT-generated trajectories with the {\it npol} and {\it pol} models are plotted in comparison to DFT results at 650 K with fixed volumes $V_0$ and $V_1$, respectively. The trajectories generated with the {\it npol} model at temperatures of 600 K and 650 K show a good agreement with the DFT results obtained for the expanded volume $V_1$ ($\Delta V$=+21\%, see Table~\ref{Volume_optimized}). This is consistent with the supercell expansion observed at finite temperature (+15\% at 600 K and +19\% at 650 K, see Supporting Information Table S4). As for the 0 K case, the stable $\alpha$ phase trajectories produced with both the {\it npol} and {\it pol} FFs accurately reproduce the DFT reference, but only the {\it npol} model can generate stable trajectories for temperatures $\geq$600 K, i.e the temperature range for which, experimentally, the $\alpha$ phase is found to be more stable than the $\delta$ phase\cite{beal_cesium_2016,marronnier_anharmonicity_2018}.  

\begin{figure}
    \centering
    \includegraphics[width=18cm]{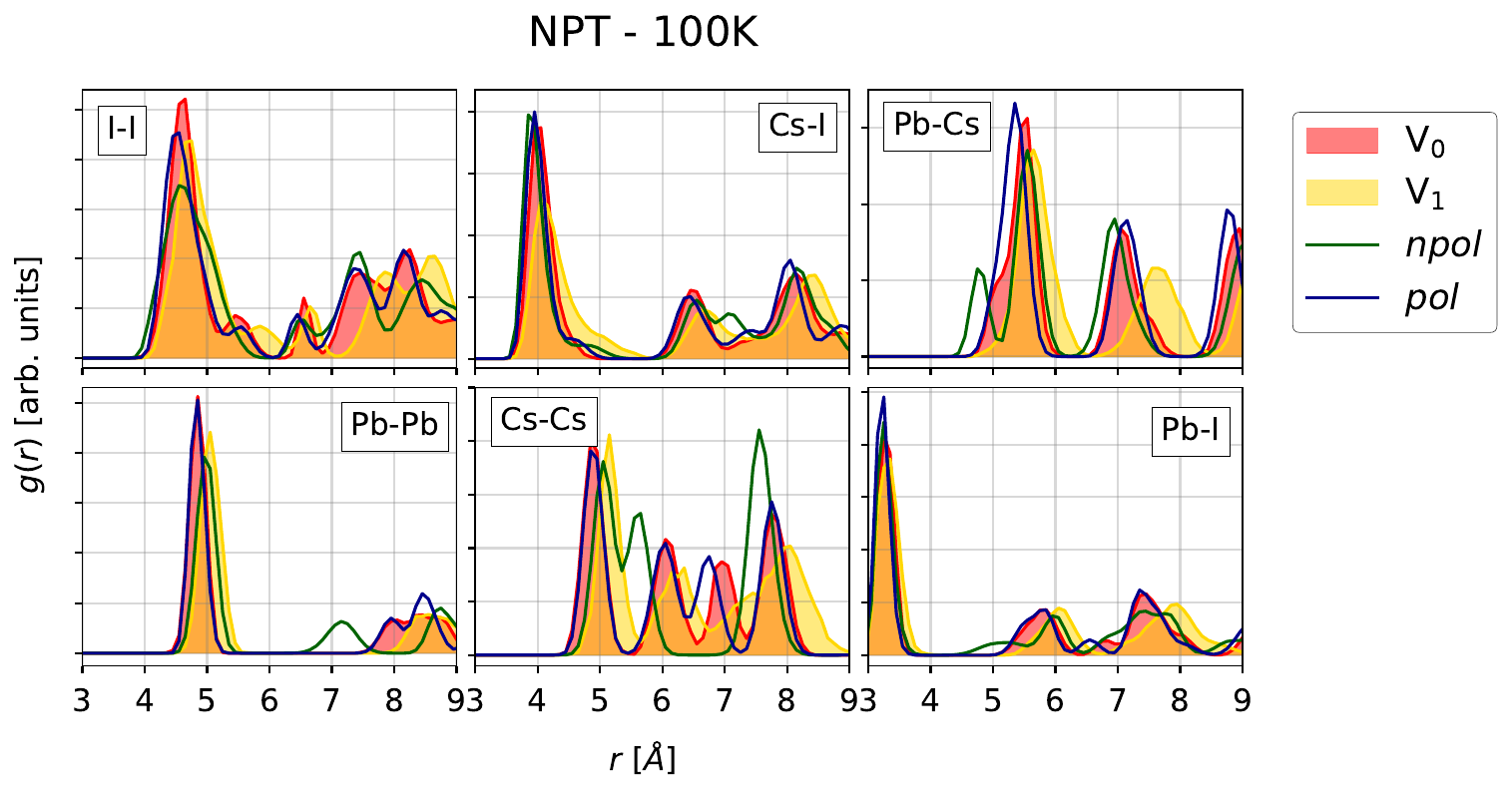}
    \caption{Comparison of radial pair distribution of DFT reference (V$_0$ and V$_1$),  {\it npol}  and  {\it pol} NPT trajectories for the $\delta$ phase. All the trajectories are those simulated at 100 K.}\label{gofr_delta_100}
    
\end{figure}

Similarly to what was observed already for the 0 K structures in Section~\ref{0Kstructuralporperties}, the results of the {\it npol} FF for the $\delta$ phase show some deviations with respect to the DFT reference. This is clear in the RPDFs at 100 K plotted in Figure~\ref{gofr_delta_100}. 
Figure~\ref{gofr_delta_100} also shows that the {\it pol}  model, on the other hand, reproduces the $\delta$ phase at 100 K in excellent agreement with respect to the DFT reference. The same is observed for higher temperatures for which the RPDFs are shown in the Supporting Information.

\subsubsection{Vibrational Properties}
\begin{figure}
    \centering
    \includegraphics[width=16cm]{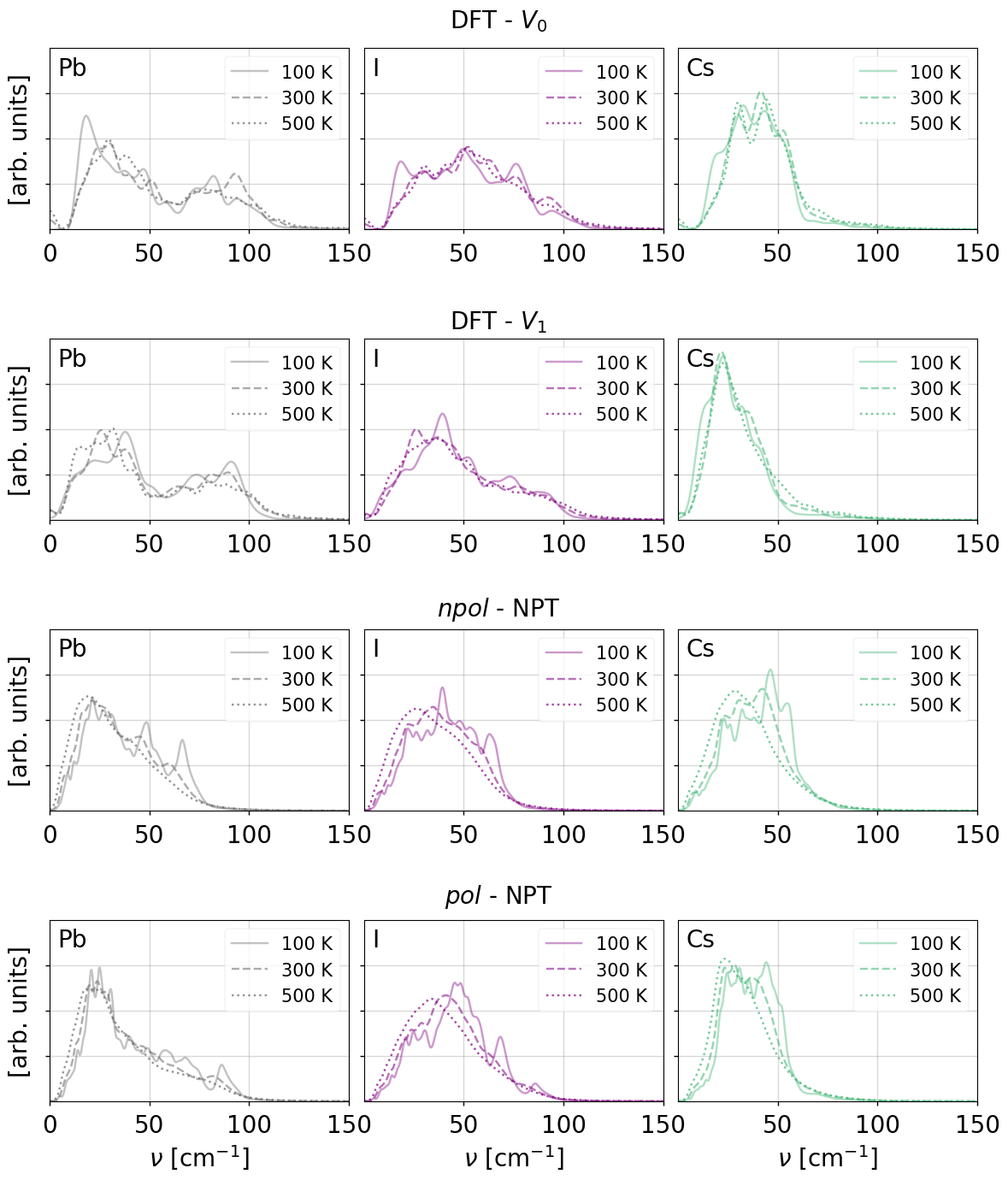}
    \caption{Species projected power spectra of $\delta$ CsPbI$_3$. The upper panels show the DFT power spectra up to 500 K for cells of volumes $V_0$ and $V_1$. The lower panels show the power spectra of NPT trajectories generated with the {\it npol} and {\it pol} FFs. Note that for the FFs model we included trajectories up to 500 K.} 
    \label{power_delta_NPT}
\end{figure}

\begin{figure}
    \centering
    \includegraphics[width=16cm]{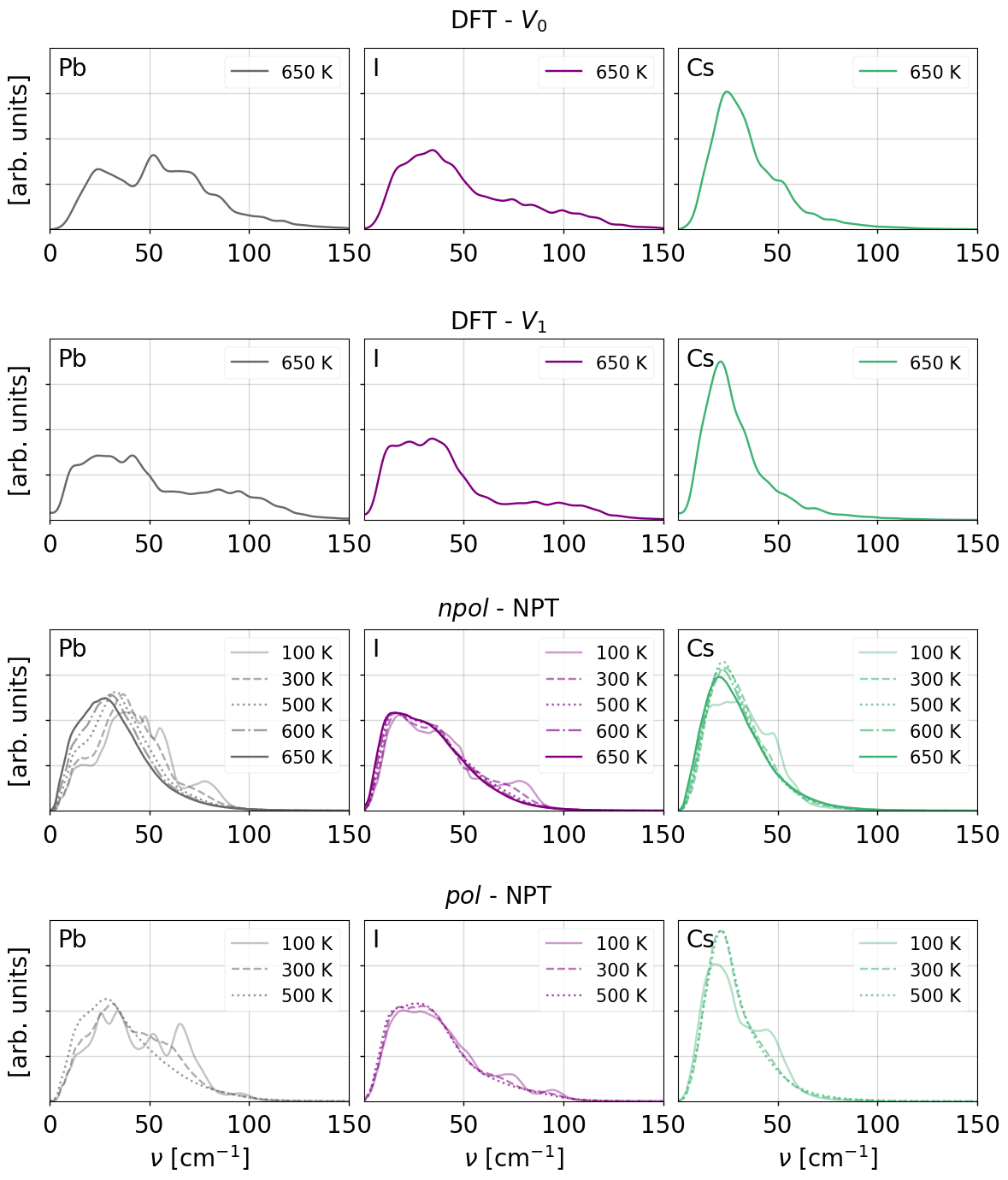}
    \caption{Species projected power spectra of $\alpha$ CsPbI$_3$. The upper panels show the DFT reference at 650 K for cell volumes $V_0$ and $V_1$. The lower panels show the power spectra of NPT trajectories generated with the {\it npol} and {\it pol} FFs. For the {\it pol} model the power spectra are shown up to 500K, which is the higher temperature non-melting trajectory. }
    \label{power_cubic_NPT}
\end{figure}

We also investigated the vibrational properties of $\alpha$ and $\delta$ phases of CsPbI$_3$ and compared the FF results with the DFT reference. In Figs.~\ref{power_delta_NPT} and~\ref{power_cubic_NPT}, we show the power spectra split into the contributions due to the individual species for $\delta$ and $\alpha$ CsPbI$_3$, respectively, for trajectories generated with different methods, temperatures and cell volumes. 
Since CsPbI$_3$ is solely constituted by heavy elements, only very soft modes, in a region of $\sim$20-120 cm$^{-1}$, are present, which is in agreement with the range typically observed for e.g. Pb-I skeleton motions in MAPbI$_3$ \cite{perez-osorio_vibrational_2015,mattoni_temperature_2016}. Comparing the $\delta$ phase DFT power spectra at different volumes, we notice that the main effect of the volume expansion is a shift of modes involving Cs to the lower frequency, probably as a direct consequence of the increase of the space in between the edge-sharing Pb-I pillars in which the Cs ions can move. The main temperature effect is the appearance of a small diffusive component for T $\geq$ 300 K, which is slightly more prominent in the $V_0$ trajectories. This is probably due to the fact  that the cell volume is kept fixed, allowing for no thermal expansion. In the $\alpha$ phase, on the other hand, volume expansion shifts the modes around 50-80 cm$^{-1}$ assigned to Pb-I stretch vibrations \cite{perez-osorio_vibrational_2015} to lower frequency. 
For higher temperatures and expanded volumes, the spectra of the two phases get very similar. Comparing the power spectra of the FF trajectories with the ones of the DFT references, some discrepancies in the higher frequency range can be observed for the modes involving Pb and I. Whereas the DFT reference spectra extend to frequencies $\geq$ 100 cm$^{-1}$, the corresponding frequency range in the FF models reduces to around 10-80 cm$^{-1}$ in the {\it npol}  and to 10-100 cm$^{-1}$ in the {\it pol} models. 
Also in this case, at high temperatures, the frequency range is shifted to a lower range and the $\delta$ and $\alpha$ phases' power spectra become more similar. Moreover, at high temperatures the differences between the results obtained with the {\it pol}, respectively {\it npol} models become smaller. In conclusion, neither of the FFs is able to fully reproduce the Pb and I modes in the higher frequency range suggesting that they are not able to fully describe the Pb-I stretch vibrations, but the {\it pol} FF including frequencies up to 100 cm$^{-1}$ provides a better match with respect to the DFT power spectra. In the Supporting Information, the species-resolved power spectra computed from the classical NVT trajectories as well as the total ones are given.

\subsubsection{Energy analysis}\label{finite_T_energy}
\begin{figure}[h!]
    \centering
    \includegraphics[width=16cm]{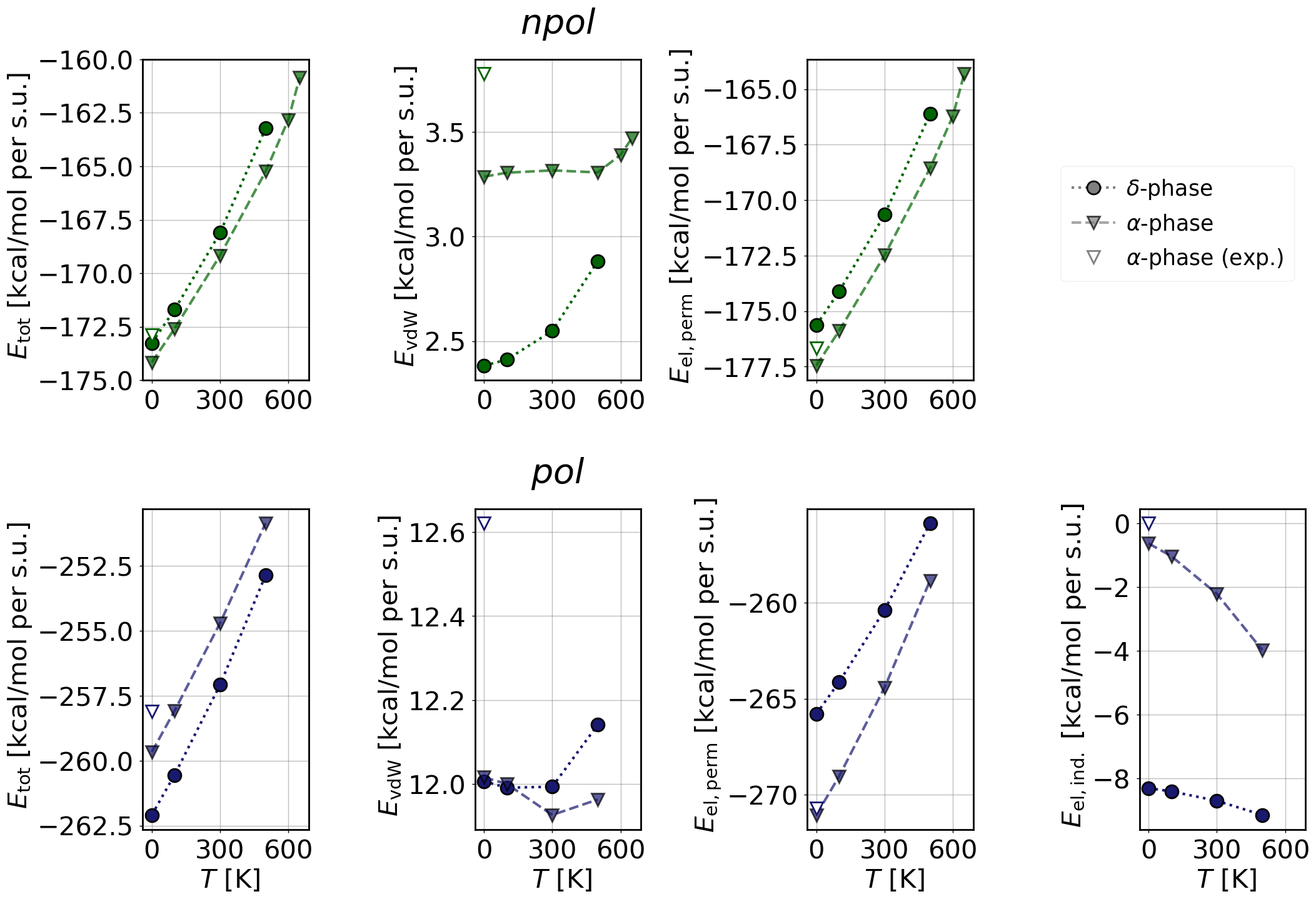}
    \caption{Average total energies and energy contributions of the classical NPT trajectories at finite temperatures of the $\delta$ and $\alpha$ phases of CsPbI$_3$. The empty points show the energy contribution of the $\alpha$ phase optimized from the experimental structure. }
    \label{Total_energy_finate_temperature}
\end{figure}

As mentioned above, the structure of the $\alpha$ phase optimized starting from the experimental (fully cubic) data ends up in a local minimum. If one considers the (orthorhombic-like) $\alpha$ phase structure optimized starting from a 100 K frame, the correct relative energetic ordering at 0 K, \emph{i.e.} the $\delta$ phase being more stable than the $\alpha$ phase, is reproduced only by the $pol$ model. Here, we study how the relative energetics of the two phases predicted by the two models change as a function of temperature. In Figure~\ref{Total_energy_finate_temperature}, the average total energy and its components from NPT trajectories at different temperatures are plotted (corresponding NVT results are given in the Supporting Information). Values of the $\alpha$ structure optimized from the experimental data are in gray to show the difference with respect to the results obtained with the 100 K optimized structure. The latter looks more consistent with the values at finite temperature. This is especially evident in the total energy and vdW terms.

In the {\it npol} model, the relative energetic order shows an $\alpha$ phase that is consistently more stable than the $\delta$ polymorph. On the contrary, in the {\it pol} model, the relative energetic ordering at low temperature is correctly reproduced and the energetic difference between the two phases steadily diminishes as a function of temperature. However, the $\alpha$ phase melts around 500K, i.e. before reaching the crossover point. 

An analysis of the different energy contributions shows that in the $npol$ model the vdW component favors the $\delta$ phase. However, differences in the vdW contribution are small in comparison to those of the electrostatic energy, which preferentially stabilizes the $\alpha$ structure. 
Also in the $pol$ model, the electrostatic interactions at finite temperature favor $\alpha$, whereas the energy differences in the vdW contribution are so small that we cannot determine for sure which phase they would favor. 
The decisive contribution to the relative energetics and its thermal evolution is by far the induced electrostatic term. Due to the inherent structural anisotropy, the $\delta$ phase has a much higher contribution of $E_{\text{ind}}$ than the $\alpha$ phase, which due to its high cubic symmetry has essentially no contributions from the induced dipole interaction at low temperature. For both phases, the magnitude of the induced electrostatic interactions becomes larger with increasing thermal fluctuations. However, due to the breaking of the initial high symmetry, the thermal increase in $E_\text{ind}$ is more rapid for the $\alpha$ leading to an overall reduction of the average energy difference between the two phases and promoting a phase transition.  
The different trend observed for the energy contributions of the {\it npol} model, in particular for the vdW term that stabilizes the $\delta$ phase, is due to the missing polarizability, which as just discussed is fundamental for obtaining the right energetic order.
   
\section{Conclusions}
Here, we presented a force matching approach for the semi-automated generation of a fixed-point charge $\it npol$ as well as for a polarizable $\it pol$
FF for the all inorganic perovskite CsPbI$_3$ based on extensive data from DFT based molecular dynamics simulations. We find that
both models are able to reproduce the overall structural features of the $\delta$ as well as the $\alpha$ phase. In particular, if one is interested in the properties of the perovskite phase alone, the $npol$ FF developed here offers a viable and computationally efficacious option. However, some subtle structural details of the highly-anisotropic $\delta$ phase are better reproduced with the $\it pol$ model, a trend that is also reflected in a better reproduction of the vibrational properties. Remarkably,    
the relative phase stability of the two polymorphs is only captured by the $\it pol$ model. In fact, it turns out that the decisive component that governs the relative phase stability is the electrostatic interactions due to induced dipoles. At lower temperatures, this contribution is stabilizing the anisotropic $\delta$ phase over the $\alpha$ phase, for which the $E_{ind}$ contribution is essentially zero due to the high symmetry. However, this contribution increases rapidly at elevated temperatures due to the symmetry breaking by thermal fluctuations. Clearly, these interactions cannot be captured with a $\it npol$ model. In fact, within the $\it npol$ model, these crucial force contributions can only be mimicked (and only partially be accounted for) by an effective vdW term. In other words, $\it npol$ force fields have to describe the relative energetic difference between $\alpha$ and $\delta$ phases based `on the wrong physics', which clearly hampers their viability and scope. The finding that the presence and magnitude of induced dipoles is the crucial factor that influences the relative energetics of $\delta$ versus $\alpha$ phase is most likely not limited to the $\delta/\alpha$ polymorphism in CsPbI$_3$ but seems also applicable to other systems such as FAPbI$_3$ and can thus provide some rational guidance for synthetic attempts in stabilizing the photoactive $\alpha$ phase. In fact, this result suggests that breaking the symmetry of the cubic phase by introducing anisotropy will help in shifting the relative phase equilibrium towards the $\alpha$ phase. This paradigm is indeed consistent with and is able to rationalize a number of experimental observations such as the fact that at low temperatures orthorhombic perovksite phases are usually more stable than the cubic ones \cite{stoumpos_semiconducting_2013}; the presence of a more anisotropic monovalent cation like FA leads to a lowering of the phase transition temperature (FAPbI$_3$ 418K \cite{sturdza_direct_2024} versus 600K for CsPbI$_3$\cite{beal_cesium_2016,marronnier_anharmonicity_2018}); and the introduction of different monovalent cations and/or other halides in mixed cation/halide perovskites can be used to stabilize the $\alpha$ phase.\cite{pellet_mixed-organic-cation_2014,saliba_cesium-containing_2016,boziki_why_2020,syzgantseva_stabilization_2017}

\section*{Acknowledgment}
C.V. gratefully acknowledges a INSPIRE Potentials - MARVEL Master's Fellowship from the NCCR MARVEL. U.R. acknowledges funding from the Swiss National Science Foundation via grant No. 200020219440 and computational resources from the Swiss National Supercomputing Centre CSCS.

\section*{Data availability} 
Data and analysis scripts are available on Zenodo at \url{https://doi.org/10.5281/zenodo.11175491}. Additionally, on Zenodo the link to the GitHub repository containing the interface for performing the force matching is provided.

\providecommand{\latin}[1]{#1}
\makeatletter
\providecommand{\doi}
  {\begingroup\let\do\@makeother\dospecials
  \catcode`\{=1 \catcode`\}=2 \doi@aux}
\providecommand{\doi@aux}[1]{\endgroup\texttt{#1}}
\makeatother
\providecommand*\mcitethebibliography{\thebibliography}
\csname @ifundefined\endcsname{endmcitethebibliography}  {\let\endmcitethebibliography\endthebibliography}{}

\end{document}